\documentstyle[twoside,graphicx]{article}

\catcode`\@=11
\long\def\@makefntext#1{
\protect\noindent \hbox to 3.2pt {\hskip-.9pt  
$^{{\eightrm\@thefnmark}}$\hfil}#1\hfill}		

\def\@makefnmark{\hbox to 0pt{$^{\@thefnmark}$\hss}}	
	
\def\ps@myheadings{\let\@mkboth\@gobbletwo
\def\@oddhead{\hbox{}
\rightmark\hfil\eightrm\thepage}   
\def\@oddfoot{}\def\@evenhead{\eightrm\thepage\hfil
\leftmark\hbox{}}\def\@evenfoot{}
\def\sectionmark##1{}\def\subsectionmark##1{}}

\oddsidemargin=\evensidemargin
\newcounter{sectionc}\newcounter{subsectionc}\newcounter{subsubsectionc}
\renewcommand{\section}[1] {\vspace{12pt}\addtocounter{sectionc}{1} 
\setcounter{subsectionc}{0}\setcounter{subsubsectionc}{0}\noindent 
	{\tenbf\thesectionc. #1}\par\vspace{5pt}}
\renewcommand{\subsection}[1] {\vspace{12pt}\addtocounter{subsectionc}{1} 
	\setcounter{subsubsectionc}{0}\noindent 
	{\bf\thesectionc.\thesubsectionc. {\kern1pt \bfit #1}}\par\vspace{5pt}}
\renewcommand{\subsubsection}[1] {\vspace{12pt}\addtocounter{subsubsectionc}{1}
	\noindent{\tenrm\thesectionc.\thesubsectionc.\thesubsubsectionc.
	{\kern1pt \tenit #1}}\par\vspace{5pt}}
\newcommand{\nonumsection}[1] {\vspace{12pt}\noindent{\tenbf #1}
	\par\vspace{5pt}}

\newcounter{appendixc}
\newcounter{subappendixc}[appendixc]
\newcounter{subsubappendixc}[subappendixc]
\renewcommand{\thesubappendixc}{\Alph{appendixc}.\arabic{subappendixc}}
\renewcommand{\thesubsubappendixc}
	{\Alph{appendixc}.\arabic{subappendixc}.\arabic{subsubappendixc}}

\renewcommand{\appendix}[1] {\vspace{12pt}
        \refstepcounter{appendixc}
        \setcounter{figure}{0}
        \setcounter{table}{0}
        \setcounter{lemma}{0}
        \setcounter{theorem}{0}
        \setcounter{corollary}{0}
        \setcounter{definition}{0}
        \setcounter{equation}{0}
        \renewcommand{\thefigure}{\Alph{appendixc}.\arabic{figure}}
        \renewcommand{\thetable}{\Alph{appendixc}.\arabic{table}}
        \renewcommand{\theappendixc}{\Alph{appendixc}}
        \renewcommand{\thelemma}{\Alph{appendixc}.\arabic{lemma}}
        \renewcommand{\thetheorem}{\Alph{appendixc}.\arabic{theorem}}
        \renewcommand{\thedefinition}{\Alph{appendixc}.\arabic{definition}}
        \renewcommand{\thecorollary}{\Alph{appendixc}.\arabic{corollary}}
        \renewcommand{\theequation}{\Alph{appendixc}.\arabic{equation}}
        \noindent{\tenbf Appendix \theappendixc #1}\par\vspace{5pt}}
\newcommand{\subappendix}[1] {\vspace{12pt}
        \refstepcounter{subappendixc}
        \noindent{\bf Appendix \thesubappendixc. {\kern1pt \bfit #1}}
	\par\vspace{5pt}}
\newcommand{\subsubappendix}[1] {\vspace{12pt}
        \refstepcounter{subsubappendixc}
        \noindent{\rm Appendix \thesubsubappendixc. {\kern1pt \tenit #1}}
	\par\vspace{5pt}}

\newcommand{\textlineskip}{\baselineskip=13pt}
\newcommand{\smalllineskip}{\baselineskip=10pt}

\def\eightcirc{
\begin{picture}(0,0)
\put(4.4,1.8){\circle{6.5}}
\end{picture}}
\def\eightcopyright{\eightcirc\kern2.7pt\hbox{\eightrm c}} 

\newcommand{\copyrightheading}[1]
	{\vspace*{-2.5cm}\smalllineskip{\flushleft
	{\footnotesize International Journal of Modern Physics A, #1}\\
	{\footnotesize $\eightcopyright$\, World Scientific Publishing
	 Company}\\
	 }}

\def\abstracts#1#2#3{{
	\centering{\begin{minipage}{4.5in}\baselineskip=10pt\footnotesize
	\parindent=0pt #1\par 
	\parindent=15pt #2\par
	\parindent=15pt #3
	\end{minipage}}\par}} 

\newcommand{\bibit}{\nineit}

\renewenvironment{thebibliography}[1]
	{\frenchspacing
	 \ninerm\baselineskip=11pt
	 \begin{list}{\arabic{enumi}.}
	{\usecounter{enumi}\setlength{\parsep}{0pt}
	 \setlength{\leftmargin 12.7pt}{\rightmargin 0pt} 
	 \setlength{\itemsep}{0pt} \settowidth
	{\labelwidth}{#1.}\sloppy}}{\end{list}}

\newcounter{itemlistc}
\newcounter{romanlistc}
\newcounter{alphlistc}
\newcounter{arabiclistc}

\newcommand{\fcaption}[1]{
        \refstepcounter{figure}
        \setbox\@tempboxa = \hbox{\footnotesize Fig.~\thefigure. #1}
        \ifdim \wd\@tempboxa > 5in
           {\begin{center}
        \parbox{5in}{\footnotesize\smalllineskip Fig.~\thefigure. #1}
            \end{center}}
        \else
             {\begin{center}
             {\footnotesize Fig.~\thefigure. #1}
              \end{center}}
        \fi}

\newcommand{\tcaption}[1]{
        \refstepcounter{table}
        \setbox\@tempboxa = \hbox{\footnotesize Table~\thetable. #1}
        \ifdim \wd\@tempboxa > 5in
           {\begin{center}
        \parbox{5in}{\footnotesize\smalllineskip Table~\thetable. #1}
            \end{center}}
        \else
             {\begin{center}
             {\footnotesize Table~\thetable. #1}
              \end{center}}
        \fi}

\def\@citex[#1]#2{\if@filesw\immediate\write\@auxout
	{\string\citation{#2}}\fi
\def\@citea{}\@cite{\@for\@citeb:=#2\do
	{\@citea\def\@citea{,}\@ifundefined
	{b@\@citeb}{{\bf ?}\@warning
	{Citation `\@citeb' on page \thepage \space undefined}}
	{\csname b@\@citeb\endcsname}}}{#1}}

\newif\if@cghi
\def\cite{\@cghitrue\@ifnextchar [{\@tempswatrue
	\@citex}{\@tempswafalse\@citex[]}}
\def\citelow{\@cghifalse\@ifnextchar [{\@tempswatrue
	\@citex}{\@tempswafalse\@citex[]}}
\def\@cite#1#2{{$\null^{#1}$\if@tempswa\typeout
	{IJCGA warning: optional citation argument 
	ignored: `#2'} \fi}}

\def\pmb#1{\setbox0=\hbox{#1}
	\kern-.025em\copy0\kern-\wd0
	\kern.05em\copy0\kern-\wd0
	\kern-.025em\raise.0433em\box0}

\def\fnt#1#2{\footnotetext{\kern-.3em
	{$^{\mbox{\scriptsize #1}}$}{#2}}}

\def\fpage#1{\begingroup
\voffset=.3in
\thispagestyle{empty}\begin{table}[b]\centerline{\footnotesize #1}
	\end{table}\endgroup}

\def\runninghead#1#2{\pagestyle{myheadings}
\markboth{{\protect\footnotesize\it{\quad #1}}\hfill}
{\hfill{\protect\footnotesize\it{#2\quad}}}}
\font\tenrm=cmr10
\font\tenit=cmti10 
\font\tenbf=cmbx10
\font\bfit=cmbxti10 at 10pt
\font\ninerm=cmr9
\font\nineit=cmti9

\font\eightrm=cmr8

\def\qed{\hbox{${\vcenter{\vbox{			
   \hrule height 0.4pt\hbox{\vrule width 0.4pt height 6pt
   \kern5pt\vrule width 0.4pt}\hrule height 0.4pt}}}$}}

\newcommand{\npb}[3]{{\bibit Nucl. Phys.} {\bf B#1} (#2) #3.}
\newcommand{\plb}[3]{{\bibit Phys. Lett.} {\bf B#1} (#2) #3.}
\newcommand{\prd}[3]{{\bibit Phys. Rev.} {\bf D#1} (#2) #3.}
\newcommand{\prl}[3]{{\bibit Phys. Rev. Lett.} {\bf #1} (#2) #3.}
\newcommand{\nima}[3]{{\bibit Nucl. Instr. Meth.} {\bf A#1} (#2) #3.}
\newcommand{\epjc}[3]{{\bibit Eur. Phys. J.} {\bf C#1} (#2) #3.}
\newcommand{\cpc}[3]{{\bibit Comput. Phys. Commun.} {\bf B#1} (#2) #3.}
\newcommand{\pn}[3]{The OPAL Collaboration, {\bibit #3}, OPAL Physics Note 
#1, #2.}

\begin{document}

\runninghead{Study of ZZ production at LEP at $\sqrt{s} = 183 - 209$ GeV} 
{Study of ZZ production at LEP at $\sqrt{s} = 183 - 209$ GeV}

\normalsize\textlineskip
\thispagestyle{empty}
\setcounter{page}{1}

\copyrightheading{}			

\vspace*{-7mm}

\begin{flushright}
OPAL Conference Report CR453 \\ 20 November 2000
\end{flushright}		

\vspace*{0.88truein}

\fpage{1}
\begin{boldmath}
\centerline{\bf STUDY OF ZZ PRODUCTION AT LEP AT $\sqrt{s} = 183 - 209$ GeV}
\end{boldmath}
\vspace*{0.37truein}
\centerline{\footnotesize GABRIELLA P\'ASZTOR\footnote{
Permanent address: KFKI Research Institute for Particle and Nuclear Physics of
the Hungarian Academy of Sciences (KFKI RMKI), Budapest, P.O.Box 49, H-1525, 
Hungary.
Supported partially by the Hungarian Foundation of Scientific Research under 
the contract number OTKA F-023259.}}
\vspace*{0.015truein}
\centerline{\footnotesize\it CERN}
\baselineskip=10pt
\centerline{\footnotesize\it Geneva 23, CH-1211, Switzerland}

\vspace*{0.21truein}
\abstracts{
A study of Z boson pair production in e$^+$e$^-$ annihilation using the OPAL 
detector at LEP is reported. 
The ZZ production cross-section is measured,
limits on anomalous ZZ$\gamma$ and ZZZ couplings are derived, and constraints 
on models of low scale quantum gravity in extra spatial dimensions are set.}
{}{}


\vspace*{1pt}\textlineskip	
\section{Introduction}	
\vspace*{-0.5pt}
\noindent
In the Standard Model (SM), the process e$^+$e$^-$ $\to$ ZZ occurs via the NC2
diagrams.
No tree level ZZZ and ZZ$\gamma$ couplings are expected, however, physics beyond 
the SM could lead to effective couplings which could then contribute to the ZZ 
production cross-section. Deviations from the SM may arise, for example, in
models with low scale quantum gravity in extra spatial dimensions 
(LSG),\cite{LSG} due to $s$-channel graviton exchange.\cite{hewett,agashe} 


In this  paper we report on measurements of the NC2 ZZ cross-section, including
the extrapolation to final states where one or both Z bosons have invariant
masses far from $m_{\mathrm{Z}}$. These measurements, together with the angular
distribution of the observed events, are then used to extract limits on ZZZ and 
ZZ$\gamma$ anomalous couplings and constraints on the fundamental scale of LSG 
models.

\section{ZZ Selections}  
\noindent 
Since 1997 the amount of data collected by the OPAL detector\cite{opal} above
the ZZ production threshold has reached an integrated luminosity of 650 
pb$^{-1}$, divided between eight center-of-mass energies as listed in Table
1.  

\begin{table}[htbp]
\tcaption{Luminosity weighted center-of-mass energies, luminosities and measured
ZZ cross-sections in the different datasets collected by July 2000. 
The last two lines show a new 
update based on the data collected by October 2000. On the measured cross
section the first error is statistical, the second one is systematic. The error
on the luminosity measurement varies slightly from energy to energy, and it
amounts approximately to 0.3\%. The precise value of the luminosity may vary
slightly from channel to channel.}
\centerline{\footnotesize\smalllineskip
\begin{tabular}{rrr}\\
\hline
$\sqrt{s}$ (GeV) & ${\cal L}$ (pb$^{-1}$) & $\sigma_{\mathrm{ZZ}}$ (pb) \\
\hline
\vspace*{0.8mm}
182.62$\pm$0.05 & 55  & $ 0.12 \ ^{+0.20}_{-0.18} \ ^{+0.03}_{-0.02}$\\
\vspace*{0.8mm}
188.63$\pm$0.04 & 178 & $ 0.80 \ ^{+0.14}_{-0.13} \ ^{+0.06}_{-0.05}$\\
\vspace*{0.8mm}
191.59$\pm$0.02 & 29  & $ 1.13 \ ^{+0.46}_{-0.39} \ ^{+0.13}_{-0.09}$\\
\vspace*{0.8mm}
195.53$\pm$0.02 & 73 & $ 1.28 \ ^{+0.27}_{-0.25} \ ^{+0.11}_{-0.08}$ \\
\vspace*{0.8mm}
199.52$\pm$0.02 & 74 & $ 1.01 \ ^{+0.25}_{-0.22} \ ^{+0.08}_{-0.06}$\\
\vspace*{0.8mm}
201.63$\pm$0.02 & 37 & $ 1.09 \ ^{+0.39}_{-0.34} \ ^{+0.11}_{-0.08}$\\
\vspace*{0.8mm}
204.8\phantom{2}$\pm$0.1\phantom{2} & 58 & $ 1.41 \ ^{+0.34}_{-0.30} \ ^{+0.12}_{-0.09}$\\
\vspace*{0.8mm}
206.6\phantom{2}$\pm$0.1\phantom{2} & 28 & $ 0.30 \ ^{+0.37}_{-0.28} \ ^{+0.07}_{-0.04}$\\
\hline 
\vspace*{0.8mm}
204.9\phantom{2}$\pm$0.1\phantom{2} & 80 & $ 1.09 \ ^{+0.27}_{-0.24} \ ^{+0.10}_{-0.08}$\\
\vspace*{0.8mm}
206.7\phantom{2}$\pm$0.1\phantom{2} & 125 & $ 1.05 \ ^{+0.21}_{-0.19} \ ^{+0.09}_{-0.07}$\\ 
\hline
\end{tabular}}
\end{table}


All ZZ topologies but $\tau^+\tau^-\nu\bar{\nu}$ and $\nu\bar{\nu}\nu\bar{\nu}$
are covered by five separate analyses which address the final
states $\ell^+\ell^-\ell^+\ell^-$, $\ell^+\ell^-\nu\bar{\nu}$,
$\mathrm{q\bar{q}}\ell^+\ell^-$, $\mathrm{q\bar{q}}\nu\bar{\nu}$ and 
$\mathrm{q\bar{q}q\bar{q}}$. In addition, in those channels where at least one
of the Z bosons decays hadronically, a b-tagging algorithm is used to identify  
$\mathrm{b\bar{b}}\ell^+\ell^-$, $\mathrm{b\bar{b}}\nu\bar{\nu}$ and
$\mathrm{b\bar{b}q\bar{q}}$ final states. The selections are described in
detail elsewhere.\cite{zz1998,zz1999,sm2000,sm2000nov} 

All selections are designed to measure pairs of on-shell Z bosons in order to
avoid background from Z$\gamma^\star$ production. The ZZ cross-section is
defined as the contribution from NC2 diagrams to the total four-fermion
cross-section. Contributions from all other four-fermion final states,
including their interference with NC2 diagrams, are considered background, and
estimated using the grc4f Monte Carlo program.\cite{grc4f}

\textheight=7.8truein

In the case of the $\ell^+\ell^-\ell^+\ell^-$ and 
$\mathrm{q\bar{q}}\ell^+\ell^-$ channels, the Z boson masses can be
reconstructed relatively cleanly with no difficult WW background present,
therefore a simple final cut on the reconstructed Z masses is sufficient.
However, the $\ell^+\ell^-\nu\bar{\nu}$, $\mathrm{q\bar{q}}\nu\bar{\nu}$ and 
$\mathrm{q\bar{q}q\bar{q}}$ final states compete with backgrounds that have
event topologies with invariant masses close to the signal region, requiring 
likelihood selections to exploit the differences between the ZZ signal and the 
mainly WW background. In the $\mathrm{q\bar{q}}\nu\bar{\nu}$
channel the We$\nu$ and the Z($n\gamma$) backgrounds are also sizable.



\section{Results} 
\noindent  
To obtain the total ZZ cross-sections and the limits on anomalous couplings and
low scale gravity, maximum log-likelihood fits are used with Poisson 
probability density convoluted with Gaussians to describe uncertainties. The
expected number of SM events ($\mu_i$) in a selection is  given by
$\mu_i \ = \ \sigma_{\mathrm{ZZ}} \ B_i \ {\cal L}_i  \ \epsilon_i \ + \ b_i$,
where
$B_i$ is the branching ratio of ZZ to the final state,
${\cal L}_i$ is the integrated luminosity,
$b_i$ is the expected background, 
$\epsilon_i$ is the ZZ efficiency estimated primarily by the YFSZZ Monte Carlo
program\cite{yfszz} and
$\sigma_{\mathrm{ZZ}}$ is the ZZ production cross-section.
The overlap between the b-tag and non-b-tag analyses is taken into account.

The fits are dominated by the large statistical errors associated with the
small number of ZZ events. Even so, systematic 
errors\cite{zz1998,zz1999,sm2000} are included in the analyses, and
conservatively it is assumed that the bulk of these are common between channels
and center-of-mass energies. 


The results of the total cross-section
measurements\cite{zz1998,zz1999,sm2000,sm2000nov} summarized in Table~1  are in
agreement with the SM predictions\cite{zzto} as shown in Figure 1(a). In order
to further check the consistency with the SM, we have allowed the branching
ratio $BR(\mathrm{Z \to b\bar{b}})$ to be a free parameter in the fit at each
center-of-mass energy. The fitted values are plotted in Figure 1(b). Combining
all energies the average value\cite{sm2000nov} is 0.200$\pm$0.033, which is 1.5
standard deviations above the LEP1 measurement\cite{pdg} of 0.151$\pm$0.001.

\begin{figure}[t!]
\centering
\includegraphics*[scale=0.29]{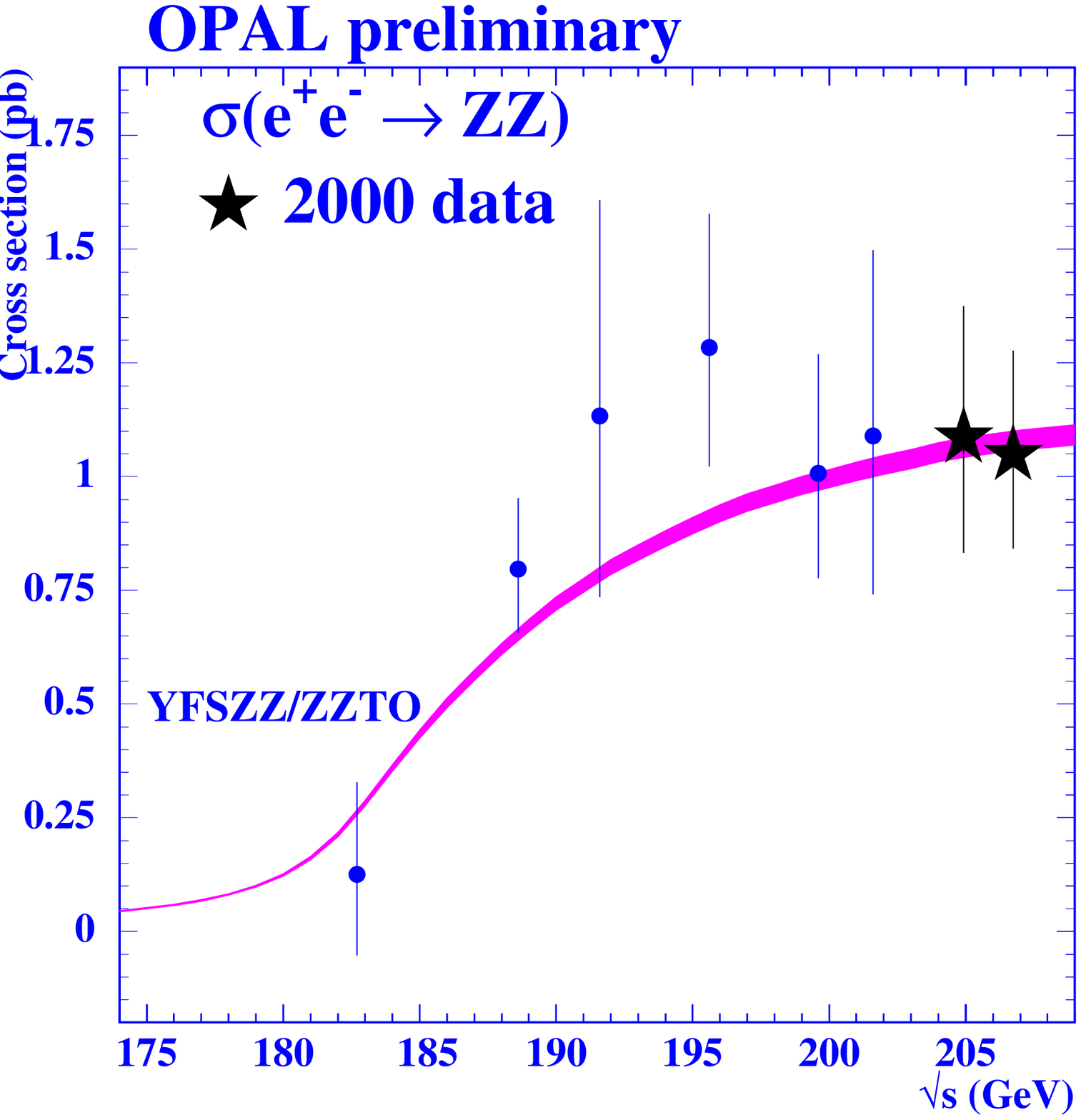} 
\includegraphics*[scale=0.29]{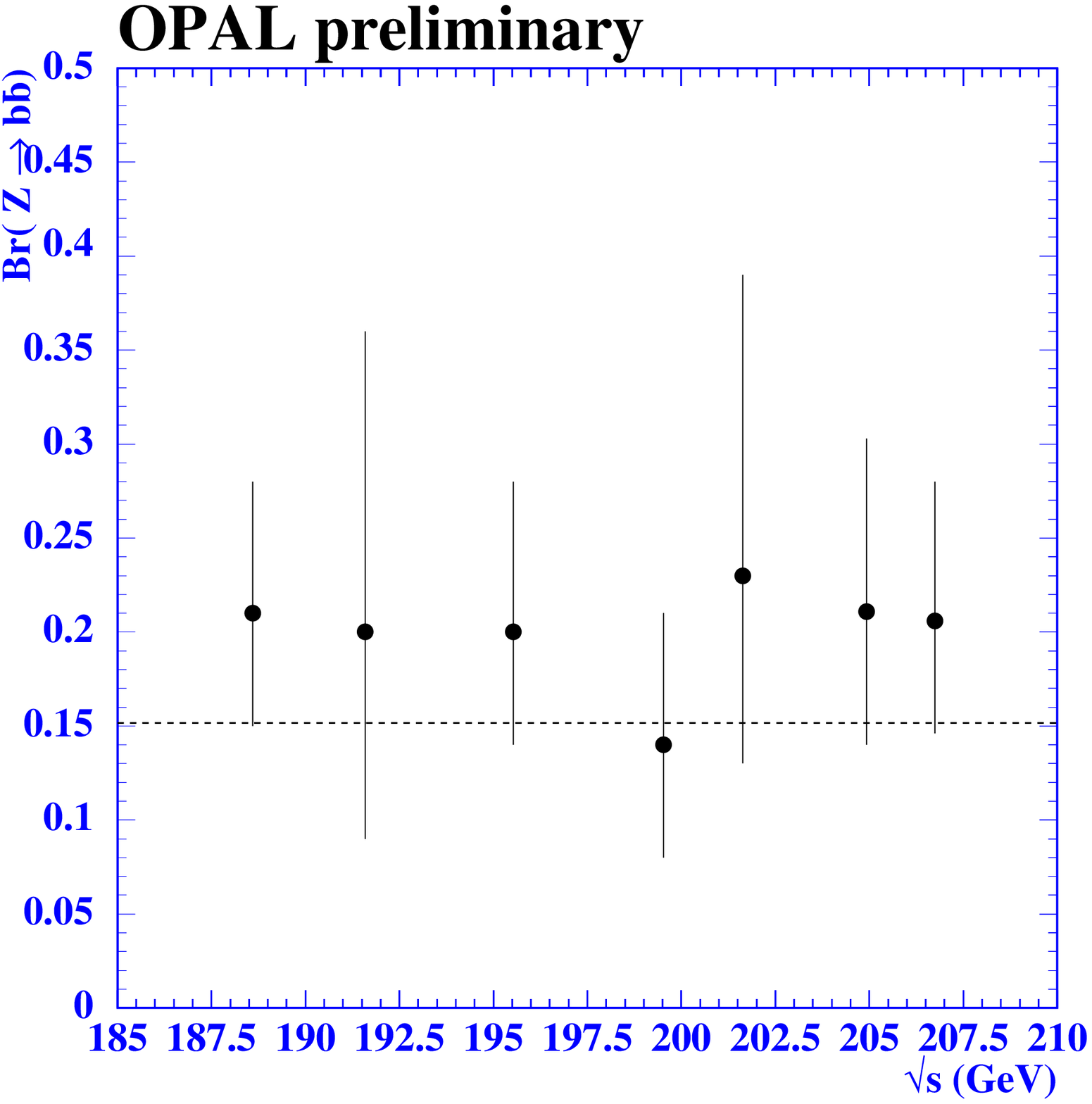} 
\vspace*{1pt}
\fcaption{(a) Measured ZZ production cross-section together with the theoretical
prediction. (b) Determination of $BR(\mathrm{Z \to b\bar{b}})$ from ZZ events
together with the measured value at LEP1. On (a) both the statistical and
systematic errors, while on (b) only the statistical errors are shown. The
common systematic error on $BR(\mathrm{Z \to b\bar{b}})$ is 5\%.}
\vspace*{-6.3cm}
\hspace*{3.8cm}(a) \hspace*{5.2cm} (b)
\vspace*{5.65cm}
\end{figure}

To derive limits on new physics we have compared the SM expectation in four
bins of $|\cos\theta|$, where $\theta$ is the polar angle of the produced 
Z boson, with our data (except at 183 GeV, where, due to the small number of
produced ZZ events, only the total rate is used). In total 327 points are used
in these fits. Additional systematic errors on the efficiency estimate are
included. To obtain these results only the datasets collected until July 2000 
are used.

Limits at the 95\% confidence level (CL) on the CP- and C-violating
$f_4^{\mathrm{VZZ}}$ and the CP-conserving, P-violating $f_5^{\mathrm{VZZ}}$
couplings, with V standing for Z or $\gamma$, are summarized in Table
2\cite{sm2000}. These are computed by fixing all other anomalous couplings to
zero and using the parametrization of Hagiwara et al.\cite{hagiwara} The limits
only slightly depend on the sign and the complex phase of couplings with the
exception of $f_5^{\mathrm{ZZZ}}$. Two-dimensional limit contours are also
derived and are shown in Figure~2.

\begin{table}[h!]
\tcaption{Limits on anomalous ZZZ and ZZ$\gamma$ couplings.}
\centerline{\footnotesize\smalllineskip
\begin{tabular}{rrr}\\
\hline
coupling & lower limit & upper limit \\
\hline
\vspace*{0.5mm}
$Re(f_4^{\mathrm{ZZZ}})$ & -0.7 & 0.8 \\
\vspace*{0.5mm}
$Re(f_5^{\mathrm{ZZZ}})$ & -1.0 & 0.5 \\
\vspace*{0.5mm}
$Re(f_4^{\mathrm{ZZ\gamma}})$ & -0.4 & 0.4 \\
\vspace*{0.5mm}
$Re(f_5^{\mathrm{ZZ\gamma}})$ & -0.9 & 0.8 \\
\hline 
\end{tabular}}
\end{table}

\begin{figure}[htbp]
\centering
\includegraphics*[scale=0.29]{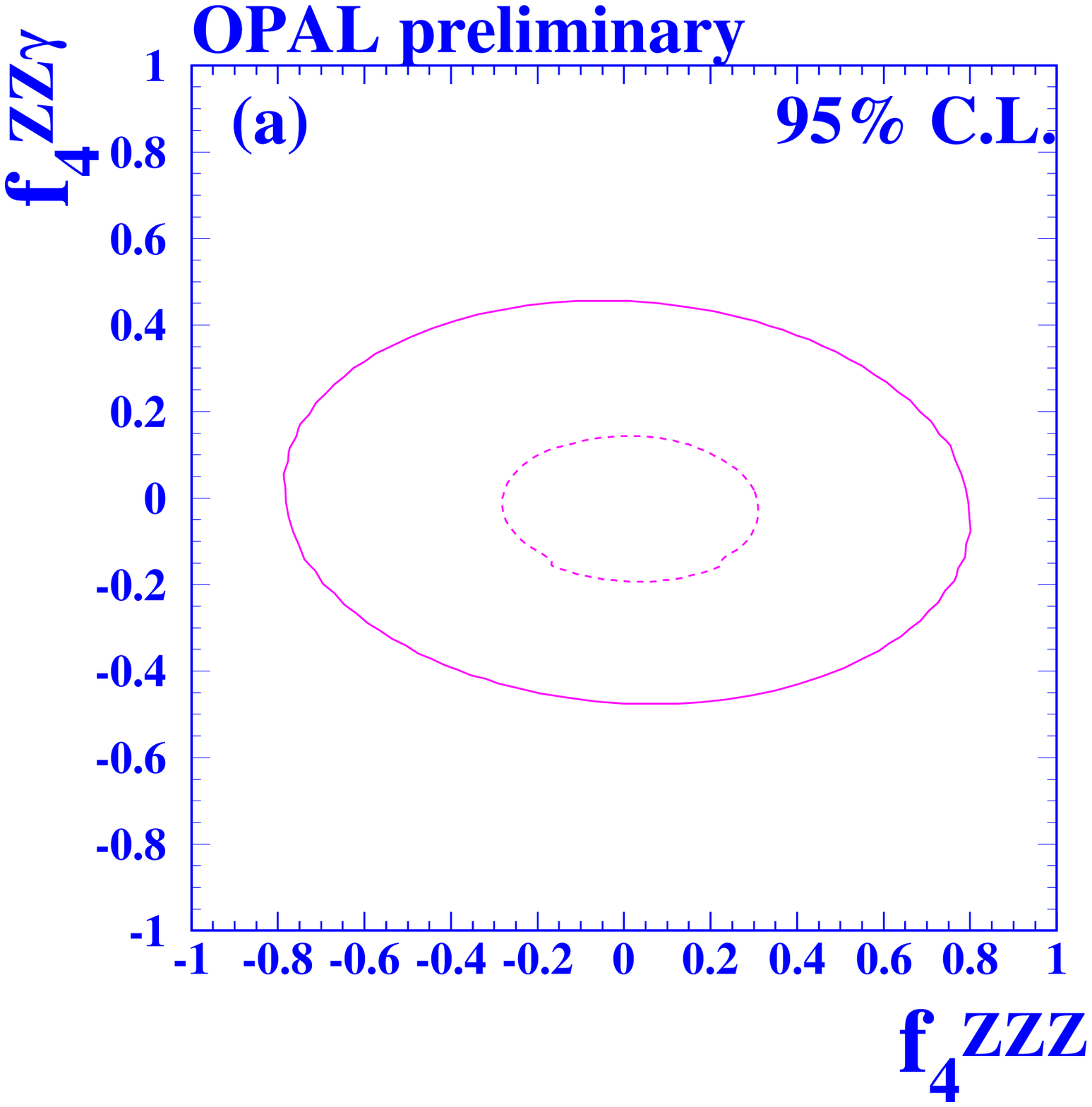} 
\includegraphics*[scale=0.29]{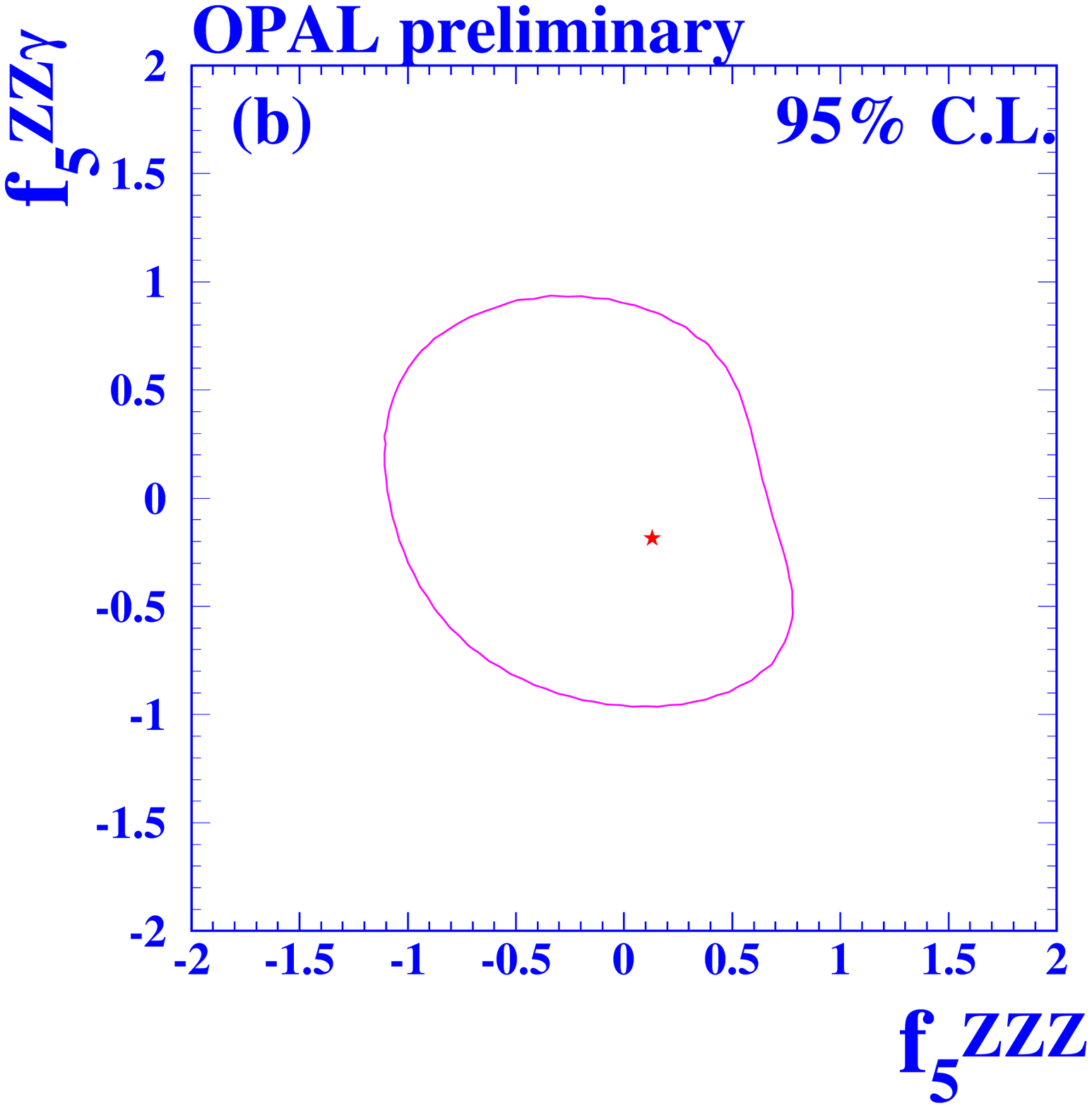}
\vspace*{-5pt}
\fcaption{Two-dimensional limit contours for anomalous couplings if they are
constrained to the (a) $f_4^{\mathrm{ZZZ}} - f_4^{\mathrm{ZZ\gamma}}$
and (b) $f_5^{\mathrm{ZZZ}} - f_5^{\mathrm{ZZ\gamma}}$ planes. In (a) the
likelihood has no unique minimum, therefore a dashed line shows the location of
the minimum values of the likelihood function. In (b) the location of
the unique minimum is denoted by a star.}
\null
\vspace*{-12pt}
\end{figure}

The Born-level matrix element in LSG models for $s$-channel graviton
exchange\cite{hewett,agashe} is proportional to $\lambda/M_S^4$. $M_S$ is an
ultraviolet cut-off parameter and is chosen to agree with the notation of
Hewett\cite{hewett}. The factor $\lambda$ incorporates any model  dependence.
The analyses of the ZZ data\cite{zz2000lsg} gives $\lambda/M_S^4 = -2.96
^{+2.77} _{-2.76}$ 1/TeV$^4$, which translates to 95\% CL lower limits on $M_S$
of 0.59 (0.83) TeV for $\lambda = -1 \ (+1)$. After the 
combination\cite{zz2000lsg}
with  OPAL studies\cite{ll1998,gg1998,gg1999,sm2000} of  e$^+$e$^-$ $\to$
$\gamma\gamma$, $\mu^+\mu^-$ and $\tau^+\tau^-$ production, the lower limits on
$M_S$ become 0.80 (0.90) TeV for $\lambda = -1 \ (+1)$.



\nonumsection{References}

\end{document}